# Carbon-substituted $MgB_2$ single crystals

Sergey Lee[1,*], Takahiko Masui[1,2], Ayako Yamamoto[1], Hiroshi Uchiyama[1,2] and Setsuko.Tajima[1]

[1] *Superconductivity Research Laboratory, ISTEC, 1-10-13 Shinonome, Koto-ku, Tokyo 135-0062 Japan*
[2] *Japan Society for the Promotion of Science (JSPS) Fellow*

Abstract

Carbon-substituted $MgB_2$ single crystals, $Mg(B_{1-x}C_x)_2$ of 0.3-1.0 mm size were grown for x=0.02-0.15 by a high-pressure technique. The doping dependence of lattice constants studied in a range of x=0.0-0.2 shows a monotonic decrease in *a*, while the *c* parameter remains almost invariant. Using X-ray diffraction and Auger electron spectroscopy, the solubility limit of C in $MgB_2$ was estimated to be about 15±1%, which is substantially larger than that reported for the polycrystalline samples synthesized by encapsulation techniques. Measurements of temperature dependence of magnetization and resistivity showed a dramatic decrease in $T_c$ with C-substitution, followed by complete suppression of superconductivity for x>0.125. Resistivity measurements in magnetic fields parallel and perpendicular to the basal plane of the crystals showed a nearly isotropic state in the heavily-doped crystals (x>0.1).

# 1. Introduction

Chemical substitution has proved to be a very powerful tool to modify the structure and physical properties of superconductors, which helps to elucidate the mechanism of superconductivity and to improve parameters important for their practical application. Although various cation substitutions in $MgB_2$ were reported during last two years after the discovery of superconductivity in this compound by Akimitsu and co-workers [1], only a few (e.a substitution of Mg by Al and Mn) were successful [2]. In contrast to the contradictory results on cation substitutions, there is a general consensus on the possibility of partial replacement of B by C in the honeycomb planes responsible for superconductivity in $MgB_2$. Formation of $Mg(B_{1-x}C_x)_2$ solid-solutions reported by several groups [3-11] showed that C substitution for B decreases the superconducting transition temperature ($T_c$) and the in-plane lattice constant (*a*), while the inter-plane lattice constant (*c*) remains almost invariant. Moreover, a positive effect of C doping on the practically important properties like $H_{c2}$, $H_{irr}$, $J_c$-T was reported recently [12,13]. At the same time, there is no consensus in the reported data on the solubility limit of C in $MgB_2$, the decreasing rate of $T_c$, and the in-plane lattice constant with carbon concentration (x). Comparing all of the reported data on the synthesis and characterization of $Mg(B_{1-x}C_x)_2$ samples, we see scattered values for the solubility limit, ranging from x=0.025 [3] to 0.15 [9] and variation in the $T_c$ of 19-35 K for a fixed x=0.1 value [7-11]. Moreover, the possibility of phase separation and formation of C-rich and C-poor $Mg(B_{1-x}C_x)_2$ phases for samples with x>0.02 was claimed based on high-resolution synchrotron X-ray powder diffraction experiments [5], but was not confirmed by neutron powder diffraction in the $Mg(B_{0.9}C_{0.1})_2$ sample [11]. The most plausible reason for contradictory results is that different synthetic routes were utilized for the preparation of C-substituted $MgB_2$ samples. Obviously, various modifications of the synthetic techniques and utilization of different materials for sealed reactors (Ta, W, stainless steel, quarts), carbon sources (elemental C or $B_4C$), atmosphere (vacuum or Ar pressure), temperature (700-1200°C) and duration of heat treatment (1-24 hrs) can produce polycrystalline samples with different amounts of C-containing impurity phases and possible inhomogeneity of C distribution in the samples.

In this paper we present the results of a systematic study of $Mg(B_{1-x}C_x)_2$ single crystals grown using a high-pressure technique and discuss the effect of C-substitution on the crystal growth, structure and superconducting properties of $MgB_2$.

## 2. Experimental

Precursors with a nominal composition of $Mg(B_{1-x}C_x)_2$ x=0.02 (C1), 0.05 (C2), 0.075 (C3), 0.1 (C4), 0.125 (C5), 0.15 (C6) and 0.20 (C7), were prepared by mechanical mixing of Mg, amorphous B and C powders. Single crystals were grown by a high-pressure technique previously developed for a pristine $MgB_2$ phase [14]. Crystal growth experiments were performed in BN containers at a pressure of 5 GPa and T=1600-1700°C using 2 step-heating to the maximal reaction temperature and further quenching to room temperature after 30 minutes of isothermal heat treatment [15]. Owing to a longitudinal temperature gradient, the temperature difference (ΔT) between the centre and outer parts of the reaction cell was estimated to be 150°C.

Single crystals of $MgB_2$ were mechanically isolated from the bulk samples and examined by a four-circle X-ray diffractometer (AFC5R, Rigaku) with $MoK_\alpha$ radiation monochromized by graphite. X-ray powder diffraction patterns were recorded using Ni filtered $CuK_\alpha$ radiation (1.5418 Å) with a MXP-18 diffractometer (MAC Science).

Crystal morphology and colour were studied using an optical microscope (Vanox-T, Olympus). The carbon content in the crystals was examined by Auger electron spectroscopy (AES) using a scanning electron microscope (JEM-7100, JEOL) operating at an accelerating voltage of 10kV. Compositional analysis was performed by a 5μm probe after 10-15 min of etching the crystal surface with an Ar$^+$ ion gun. The superconducting properties of single crystals were studied by dc magnetization using a SQUID (MPMS XL, Quantum Design) and four-probe resistivity measurements in various magnetic fields.

## 3. Results and Discussion

Figure 1 shows a typical picture of isolated C-substituted MgB$_2$ crystals. The size of the plate- and bar-like crystals varied by 0.3-1.0 mm in the (*ab*)-plane (Fig.1a) with a typical thickness along the *c*-axis of 0.1-0.3 mm. Optical microscopy observations of the C-substituted crystals showed that the (*ab*)-face of each crystal changes colour from silverish to brown for normal reflected and cross-polarized light respectively, while the (*ac*)-face changes from cooperish to blue (Fig.1b). Note that the *(ab)*-face of pristine MgB$_2$ crystal changes from silverish to grey and the *(ac)*– one from golden to blue as the polarization angle increases [16].

Good-quality single crystals were grown in all batches, except the one with the highest carbon content (C7), where the amount of impurity phases increases drastically and a bulk sample consists of rather small, randomly packed crystallites. We failed to isolate any single crystals of C7 with a size and quality sufficient for single-crystal study, thus the lattice constants of sample with a nominal composition of Mg(B$_{0.8}$C$_{0.2}$)$_2$ were determined from multiphase powder, containing impurity phases (MgB$_2$C$_2$, BN, MgO).

In Table 1, we summarize the most important experimental results obtained in this study of C1-C7 samples, including lattice constants *a* and *c* determined by XRD, T$_c$ values determined by the onset of magnetization, in-plane residual resistivity ρ$_{ab}$ and residual resistance ratios (RRR) estimated by transport measurements and relative intensities of the C peak I$_{(C)}$ in the Auger spectra. The C-substituted crystals are compared to the pristine MgB$_2$ crystals, which were obtained by rapid quenching (A) and slow cooling (B) from the reaction temperature as reported previously [15].

In Figure 2, we present the variation of lattice constants *a, c* vs C content (x) in the nominal composition $Mg(B_{1-x}C_x)_2$. With C substitution, *a* decreases while *c* remains almost constants, which is in general agreement with previous studies [4-11]. However, we found that the rate of decrease of the *a* parameter with C-doping is faster than that reported for the polycrystalline samples, while the *a* values for x 0.1 are the lowest yet reported suggesting that these crystals have the largest C concentration. Using our XRD data and assuming a linear relation of lattice constant *a* on x, the solubility limit of C in $MgB_2$ at a pressure of 5 GPa and T~1600$^o$C was estimated to be around x=0.15(1), which nearly corresponds to the substitution of one B atom by C in the hexagonal unit cell (x=1/6).

To estimate the relation between nominal composition of precursor and carbon content in the single crystals we used Auger electron spectroscopy (AES), which is extremely sensitive for detection of light elements (e.a. B and C). Since probing depth of AES is very small and the surface of $MgB_2$ is quite sensitive to degradation, reliable experimental data was obtained only after etching the crystal surface with an $Ar^+$ ion beam within the microscope chamber. Preliminary analysis showed that etching the crystals for 10-15 min completely removes surface contaminants, while further etching does not change the concentration of elements. Typical Auger spectra for C-containing $MgB_2$ crystals recorded after 15 min of $Ar^+$ etching is shown in Fig.3a. Analysis in 5 different points of randomly selected crystals by 5μm size microprobe showed excellent reproducibility of experimental data with a deviation of <1.5% from the average value of relative intensity for C-peak [$I_{(C)}/I_{(B+C)}$]. In Fig.3b, the results of AES analysis for the whole series of studied samples are presented. The dependence of relative intensity of the C-peak in the Auger spectra on C-concentration in the nominal composition of precursor shows linear behavior for all samples except C7, which has the highest carbon content. For this sample, the carbon concentration is reduced in the bulk of the crystallites, while the C-peak intensity is drastically enhanced at the surface, which can be attributed to the formation of C-rich impurity phases (e.a. $MgB_2C_2$) at the crystal surface. Extrapolating the experimental data for the lower C-content samples by linear dependence, the actual carbon content for C7 is about 15%, in excellent agreement with the solubility limit of C in $MgB_2$ determined from XRD data.

Figure 4a shows normalized dc magnetization vs temperature in zero-field cooled experiments at 10 Oe for randomly oriented pristine $MgB_2$ crystals (A) and $Mg(B_{1-x}C_x)_2$ crystals with x=0.02–0.125 (C1-C5) measured with a SQUID. All of the samples show rather sharp single-step transitions, which systematically shifts to the low-temperature region with increasing C-content (x). Note that samples C6 and C7 did not exhibit superconductivity down to 1.5K. The variation of superconducting transition temperature ($T_c$), defined by onset of magnetization, is shown in Fig.4b. One can see the nonlinear x-dependence of $T_c$. Comparing the rates of $T_c$ suppression with the change in lattice volume ($V/V_o$) by C-substitution and by applying hydrostatic pressure [17] (inset Fig.4b), we can conclude that the effect of lattice contraction on $T_c$ suppression by C-substituted $MgB_2$ is quite modest. Therefore, as a source of $T_c$ suppression by C-substitution, we need to consider some additional effect such as the electron doping, as predicted by band structure calculations [18].

Figure 5a shows a normalized resistance for the C-substituted single crystals (C1-C5) and pristine $MgB_2$ crystals (A and B) to emphasize a systematic change in the residual resistivity ratio (RRR). The absolute values of residual resistance [$R(T_c)$] were also estimated (Table 1). One can see an increase in the residual resistance from $1\mu\Omega$cm for undoped crystal (A) to 50 $\mu\Omega$cm for C5 crystal implying the increase of the impurity scattering contribution to the resistivity. The transition width determined for the crystal C5 exhibiting superconductivity at 2.5K (Fig.5b) is still very sharp <0.3K. The temperature dependence of resistivity for this crystal in magnetic fields of H//c and H//ab showed an almost isotropic electronic state. This indicates that the effect of C-substitution is not only electron-doping, but also a radical modification of the electronic structure. The loss of anisotropy in $H_{c2}$ seems to suggest that contribution of the 2-dimentional $\sigma$-bands to the Fermi surface disappears in the heavily C-doped $MgB_2$. The results of systematic single crystal study of C-doping effect on the superconducting and normal state properties of $MgB_2$ will be reported elsewhere [19].

## 4. Conclusions

In summary, single crystals of $Mg(B_{1-x}C_x)_2$ were grown by a high-pressure technique (P=5 GPa and T>1600°C) for a wide range of substitution of B by C x=0.02-0.15. The solubility limit of C in $MgB_2$ for this synthesis method was estimated to be x=0.15(1) by studying the relationship between nominal C-concentration and changes in the lattice constant *a*. The same value for the solubility limit was obtained by studying a compositional dependence of C-content by Auger electron spectroscopy. The observed rates of in-plane lattice parameter contraction and $T_c$ depression vs C-substitution are the highest yet reported. For the first time, the possibility of complete suppression of superconductivity and formation of an almost isotropic electronic state in the heavily C-doped $MgB_2$ was shown.

## Acknowledgements

This work is supported by the New Energy and Industrial Technology Development Organization (NEDO) as Collaborative Research and Development of Fundamental Technologies for Superconductivity Applications.

**Figure captions**

1. (a) A typical picture of plate- and bar-like C-substituted MgB$_2$ single crystals selected for measurements of magnetic and transport properties. Tick marks correspond to 1 mm.; (b) optical microscopy images of (*ab*)- and (*ac*)-face of C-substituted crystals (left and right side respectively) in normal reflected (top) and cross-polarized (bottom) light.

2. Lattice parameters *a* and *c* as a function of the nominal C content *x* in Mg(B$_{1-x}$C$_x$)$_2$.

3. (a) A typical Auger spectra of C-containing MgB$_2$ single crystals after etching the surface by Ar$^+$; (b) Average value of relative intensity of C peak [I$_{(C)}$/I$_{(B+C)}$] in the Auger spectra as a function of nominal C content *x* in Mg(B$_{1-x}$C$_x$)$_2$.

4. (a) Normalized zero-field magnetization measured at 10 Oe for the randomly oriented Mg(B$_{1-x}$C$_x$)$_2$ single crystals with a different C content *x*; (b) T$_c$ value determined by onset of magnetization vs nominal C content *x* in Mg(B$_{1-x}$C$_x$)$_2$. Inset show dependence of the T$_c$ on relative volume for C-substituted crystals in comparison to the data obtained for pristine MgB$_2$ (crystal B) produced in the hydrostatic pressure experiment [ref.17].

5. (a) Temperature dependent normalized resistance for the Mg(B$_{1-x}$C$_x$)$_2$ single crystals C1-C5 in comparison to pristine crystals A and B [ref.15]; (b) Temperature dependence of resistivity for crystal C5 in a zero magnetic field (cross), 100 and 500 G applied parallel (open symbols) and perpendicular (filled symbols) to the crystal *c*-axis.

Table 1 Nominal composition, lattice constants, critical temperature, resistivity, residual resistivity ratio (RRR) and relative intensity of carbon peak (in the Auger spectra) for the C-substituted $MgB_2$ and pristine [ref.15] crystals.

| $Mg(B_{1-x}C_x)_2$ x | $a$ (Å) | $c$ (Å) | $T_c$ (K) | $\rho_{ab}(T_c)$ (μΩcm) | RRR $\rho(300)/\rho(T_c)$ | $I_{[C]}/I_{[C+B]}$ (%) |
|---|---|---|---|---|---|---|
| C1(0.02) | 3.0776(5) | 3.5217(5) | 35.5 | 3 | 2.2 | 2.9(0.5) |
| C2(0.05) | 3.0641(7) | 3.5226(7) | 30 | 15 | 1.7 | 6.7(0.7) |
| C3(0.075) | 3.0524(8) | 3.5224(8) | 24 | 20 | 1.4 | 9.7(0.7) |
| C4(0.1) | 3.0449(8) | 3.5215(8) | 13.5 | 25 | 1.3 | 12.2(0.8) |
| C5(0.125) | 3.0356(6) | 3.5209(7) | 2.5 | 50 | 1.2 | 15.2(0.9) |
| C6(0.15) | 3.0322(5) | 3.5213(7) | no | - | - | 17.2(1.1) |
| C7(0.2) | 3.0236(11)* | 3.5223(14)* | no | - | - | 18.3(1.2) |
| A(0.0) | 3.0852(8) | 3.5202(8) | 38.0 | 1 | 5.5 | 0.7(0.5) |
| B(0.0) | 3.0877(5) | 3.5214(6) | 38.5 | - | 7.5 | - |

*lattice constants for C7 were determined by powder XRD

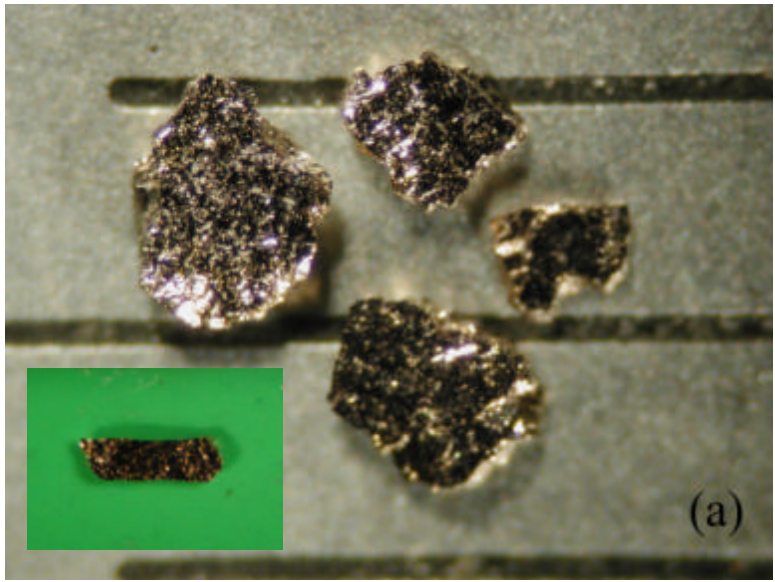

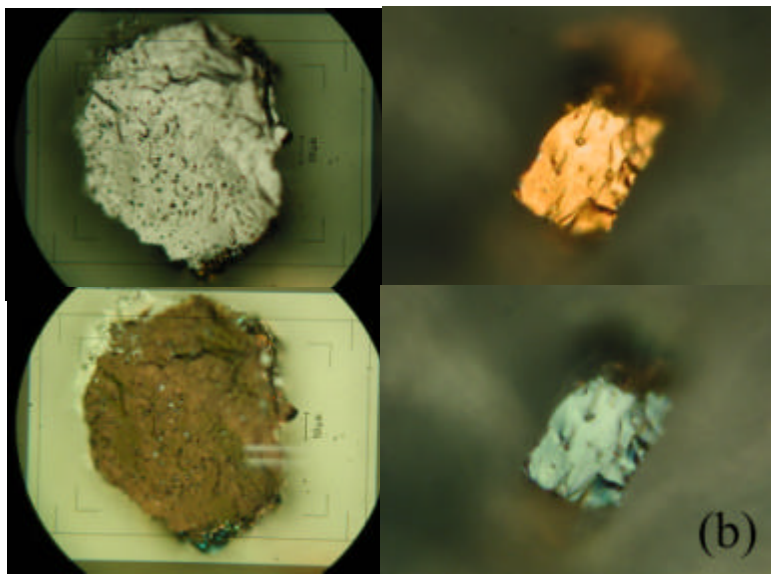

Figure 1 (S.Lee e.a.)

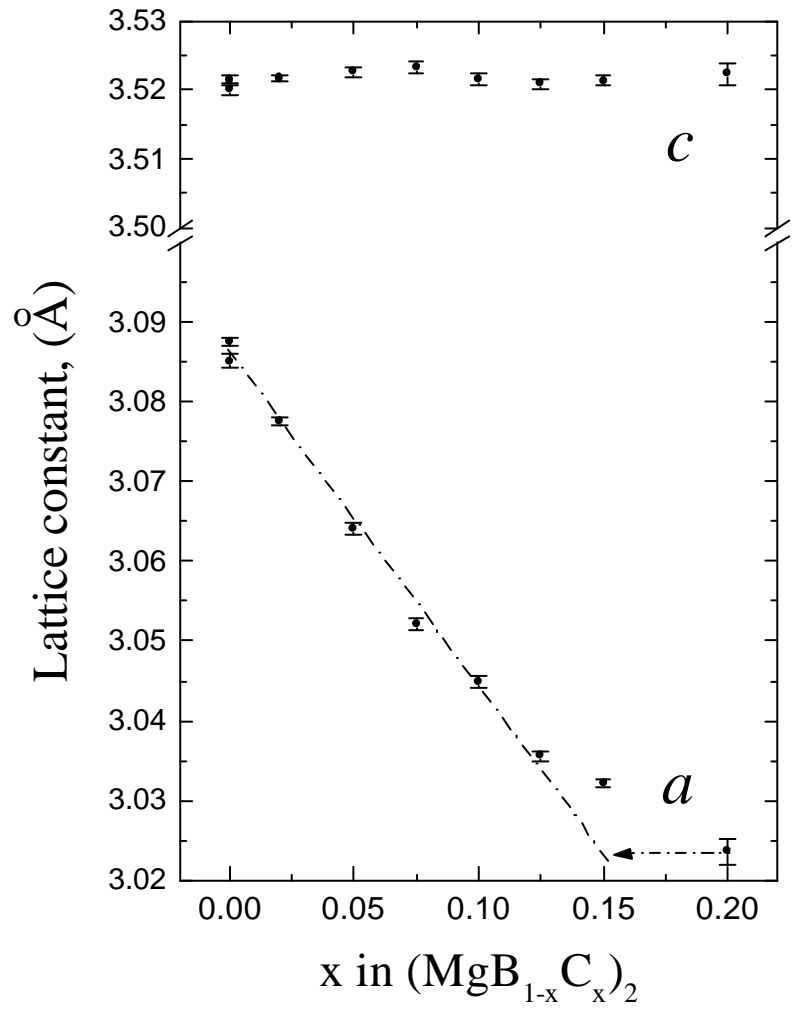

Figure 2 (S.Lee e.a.)

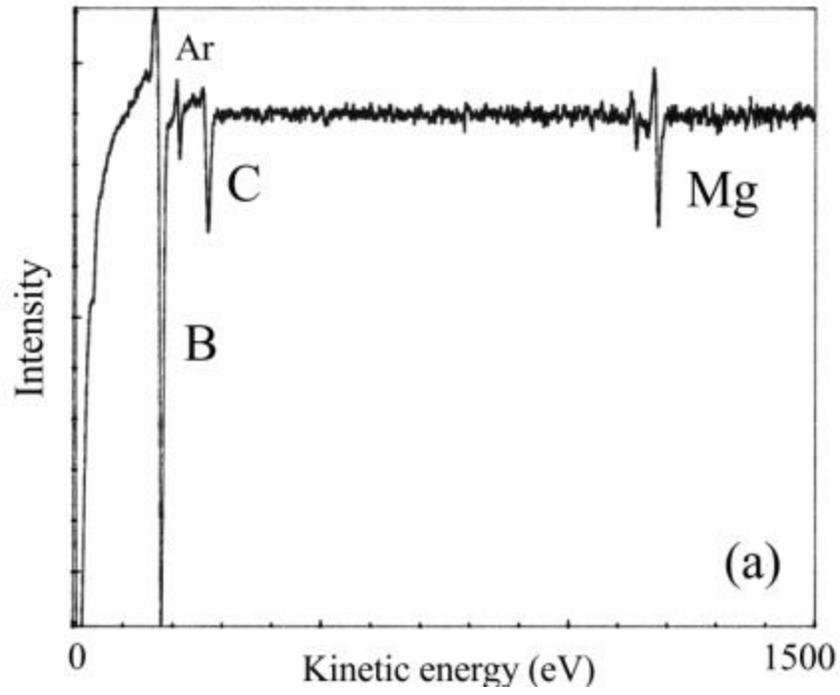

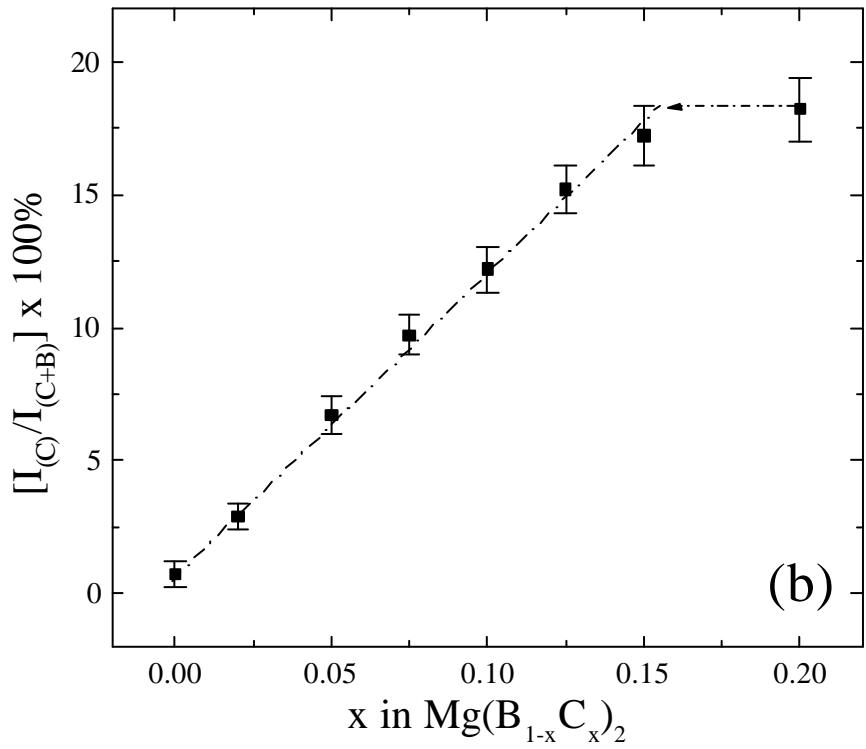

Figure 3 (S.Lee e.a.)

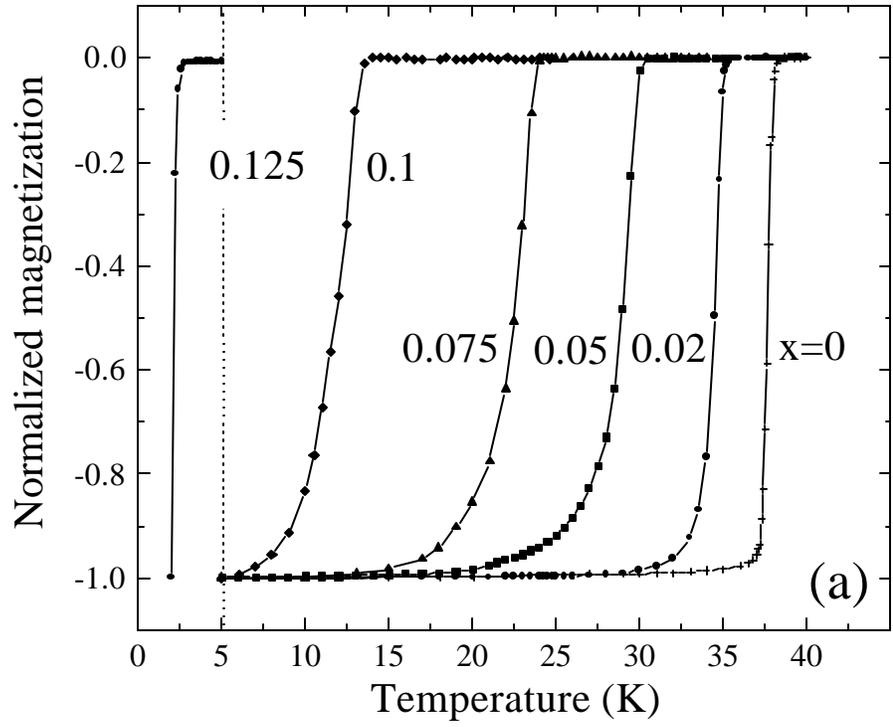

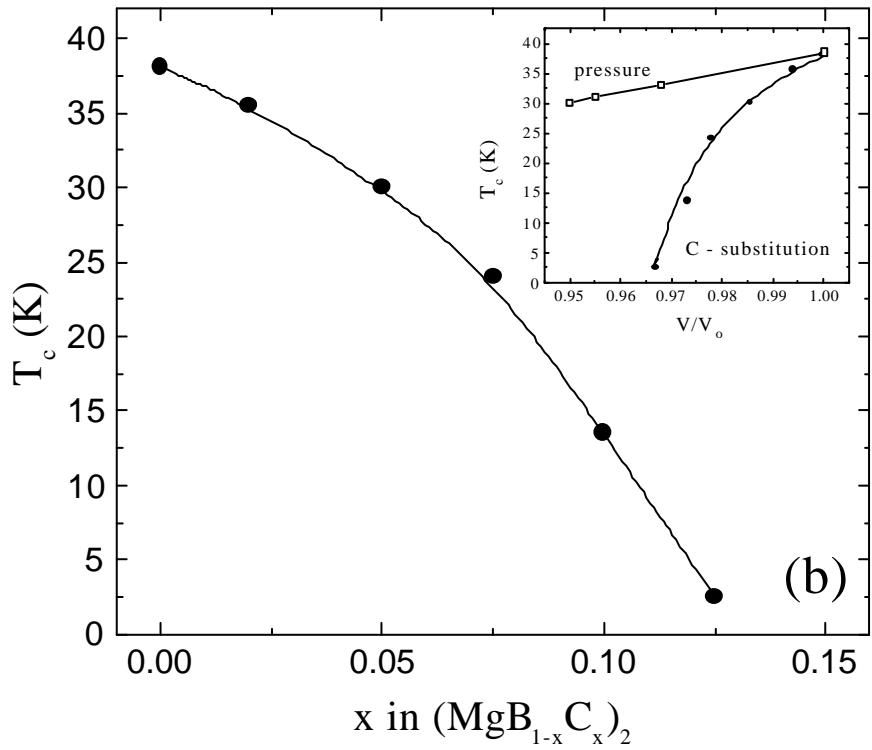

Figure 4 (S.Lee e.a.)

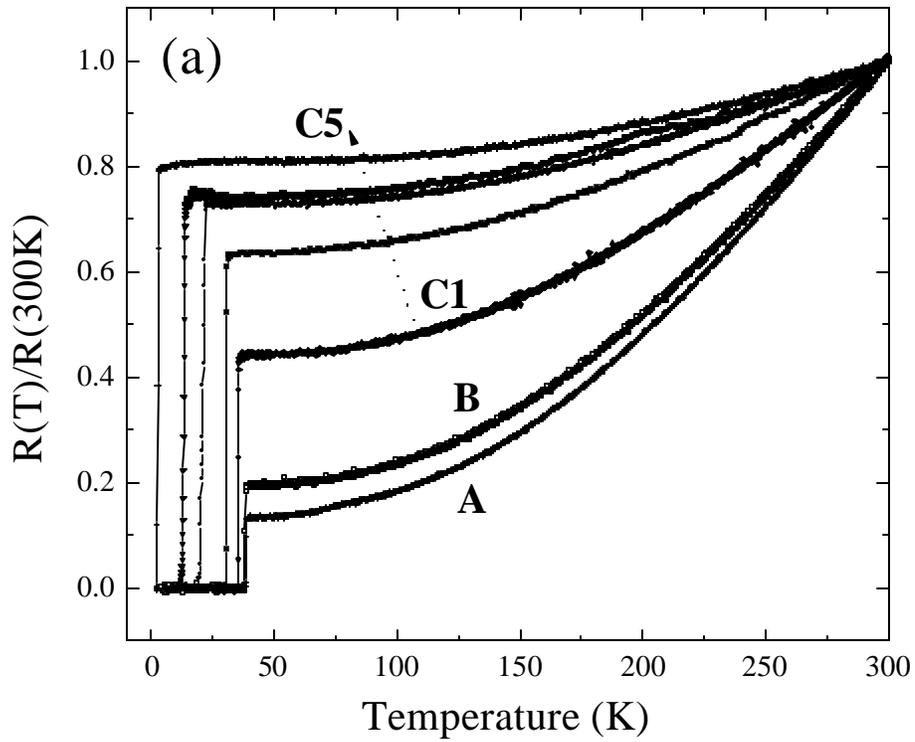
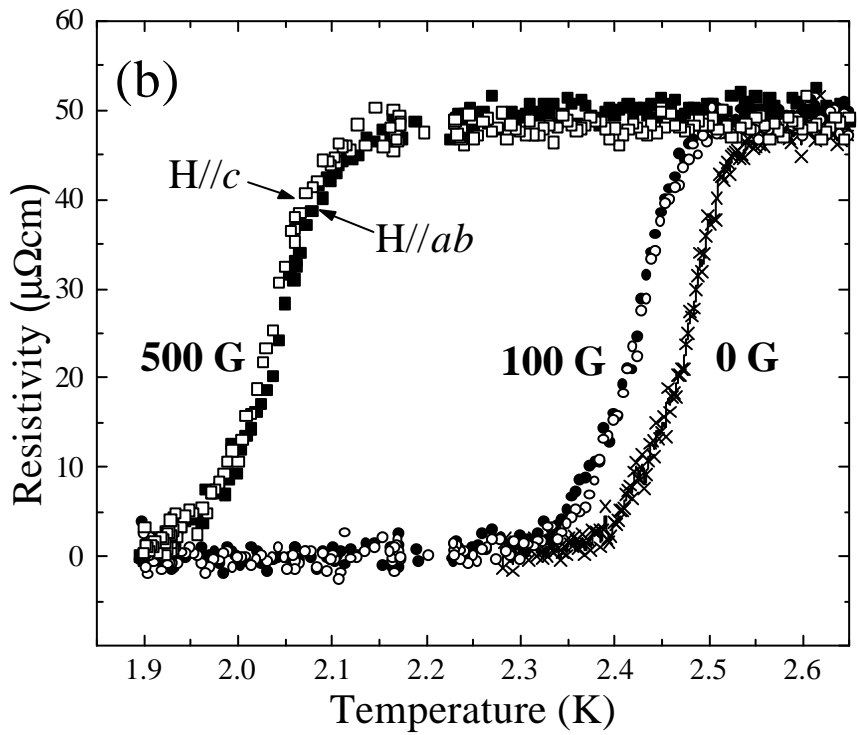

Figure 5 (S.Lee e.a.)